# COVID-19 – a realistic model for saturation, growth and decay of the India specific disease

V K Jindal[1]

Department of Physics, Panjab University, Chandigarh-160014, India

**This work presents a simple and realistic approach to handle the available data of COVID-19 patients in India and to forecast the scenario. The model proposed is based on the available facts like the onset of lockdown (as announced by the Government on 25th day, $\tau_0$ and the recovery pattern dictated by a mean life recovery time of $\tau_1$ (normally said to be around 14 days). The data of infected COVID-19 patients from March 2, to April 16, 2020 has been used to fit the evolution of infected, recovery and death counts. A slow rising exponential growth, with $R_0$ close to 1/6, is found to represent the infected counts indicating almost a linear rise. The rest of growth, saturation and decay of data is comprehensibly modelled by incorporating lockdown time controlled $R_0$, having a normal error function like behaviour decaying to zero in some time frame of $\tau_2$. The recovery mean life time $\tau_1$ dictates the peak and decay. The results predicted for coming days are interesting and optimistic. The introduced time constants based on experimental data for both the recovery rate as well as for determining the time span of activity of $R_0$ after the lockdown are subject of debate and provide possibility to introduce trigger factors to alter these to be more suited to the model. The model can be extended to other communities with their own $R_0$ and recovery time parameters.**

INTRODUCTION

While the world was entering into the new leap year 2020 with new objectives and targets, Corona Virus took control over the world rather stealthy but quickly and strongly. Waking up lethargically in the new year, we slowly had to accept the reality that most part this year was an exceptional year of destruction without a blast. The new disease initiated during the fag end of 2019, identified as COVID-19 virus infection was declared Pandemic by WHO on March 11, 2020 as it engulfed almost the entire globe in a span of a couple of months. Every day the data throws more and more deaths, disease and helplessness with little hope of recovery. The scale of death, misery, helplessness and negativity caused by the disease is monumental. There is great deal of pessimism at this point of time. We don't seem to know when and if this will end and how. Therefore, during such times a convincing forecast based on acceptable physical, data based arguments becomes necessary. This paper presents simple physical model based on data to show how and when we will come over this pessimism. Though the model proposed is India specific, but in quick time the results can be recalculated for any country or community using the same source code.

[1] Presently Honorary Professor at Department of Bio and Nano Sciences, Guru Jambheshwar University Hisar -India, email vkjindal06@gmail.com

Ever since a study on COVID 19 for India updates by Eili Klein et al[1] appeared on March 24, 2020, representing monumental scale of devastation in India through numbers of people, there has been a great concern how to prepare for such a massive possible tragedy. The alarming number of deaths predicted in this work motivated me to apply simple physical methods of growth and decay to make a fresh assessment. Indeed simulation models many times lack in insight and non-transparent making it difficult to find the control factors of the real processes.

Recently, another interesting study by Matjaz et al [2] has present outlook for the United States, Slovenia, Iran, and Germany. They show that the epidemic growth is a highly non-linear process in the sense it matters what actions are taken by the Governments and so on. It seems they dwell mostly on the available data for their forecasts. They assume recovery time and mortality % for their predictions. There are other studies on similar lines [3-10] expressing projections of this epidemic, including some theoretical analysis of epidemic sometimes using statistical physics have been made in the past [11-15]. However, in our view, there is dearth of research work which addresses simple issues and resolve them with transparent logic.

It is clear that we have sufficient data of the epidemic growth of various communities and countries. Most of this data shows exponential growth with a reproductive rate of growth denoted by $R_0$. The rate factor $R_0$ has been found to be variable depending upon community and Country. The other factors which can effectively reduce $R_0$ are lockdown and social distancing and use of masks etc. The dynamics of $R_0$ is one important key factors of this paper. The other factors are the onset of lockdown and the rate of recovery. The 3 factors which control the flattening and decay of the evolution of infected number of persons are judiciously described based on input and physical arguments.

**THEORETICAL PROCEDURE**

Since most of the data and details are well understood, therefore I straight away come to the following theoretical procedure.

The basic eq. that dictates the number of infected persons $N_e(t)$ at a time t in an environment which has a total population of N in an area A, assumed to have an average uniform number density $\sigma_0 = N/A$ is given by

$$\frac{dN_e}{dt} = \left(\frac{dN_e}{dt}\right)_{R_0} - \left(\frac{dN_e}{dt}\right)_{\tau} \qquad (1)$$

Where the net rate of increase of $N_e(t)$ is decided by the increase through reproductive rate number $R_0$ and decrease by a 'relaxation time' $\tau$, which tries to restore the deviation in some mean time $\tau$.

The first term in Eq.(1) depends on growth rate R and is given by

$$\left(\frac{dN_e}{dt}\right)_{R_0} = RN_e, \qquad (2)$$

One important criterion which I introduce is that R is not a constant but depends on time, e.g under lockdown conditions as and when imposed, R will tend to vanish in some time.

Similarly, the decay factor is like out of balance factor controlled by a recovery mean time,

$$\left(\frac{dN_e}{dt}\right)_\tau = \frac{N_e}{\tau} \tag{3}$$

Thus combining these, the eq. 1 reduces to

$$\frac{dN_e}{dt} = (R - 1/\tau)N_e \tag{4}$$

This results in

$$N_e(t) = N_e(0)e^{(\int R(t)dt - t/\tau)} \tag{5}$$

In eq. 5, the argument of exponential has to stay under an integral sign as it has not been assumed constant. In our analysis, we assume R is a function surface number density σ which keeps on fluctuating with time depending upon the movement of the number of people. For coronavirus it is stated to be R=0, for σ≤1/4, i.e. when number of people separate themselves by social distance of ≥ 2 meters. The number density keeps on fluctuating randomly if the persons in the area are not static.

The fluctuations in σ are statistical in nature and cause R to vary even under lockdown conditions. We assume therefore it legitimate to treat these variations to be like a Gaussian or an error function in nature for t greater than $t_0$.

$$R(t) = R_0 e^{-\alpha(t-t_0)^2} \tag{6}$$

where $t_0$ is the time when lockdown is initiated, and α determines how quickly R goes to zero. In our calculations, R is considered to be controlled by some time constant $\tau_2$ assumed to be close to the mean life time of virus, i.e. $\tau_1$. However, it has to be adequately adjusted if the conditions of lockdown are getting violated.

$$\alpha = 1/\tau_2^2$$

The choice of α is therefore debateable and it can be profitably tailored to represent observed behaviour. A typical time dependence of R is shown in Fig.1., where $\tau_2 = 14$ days.

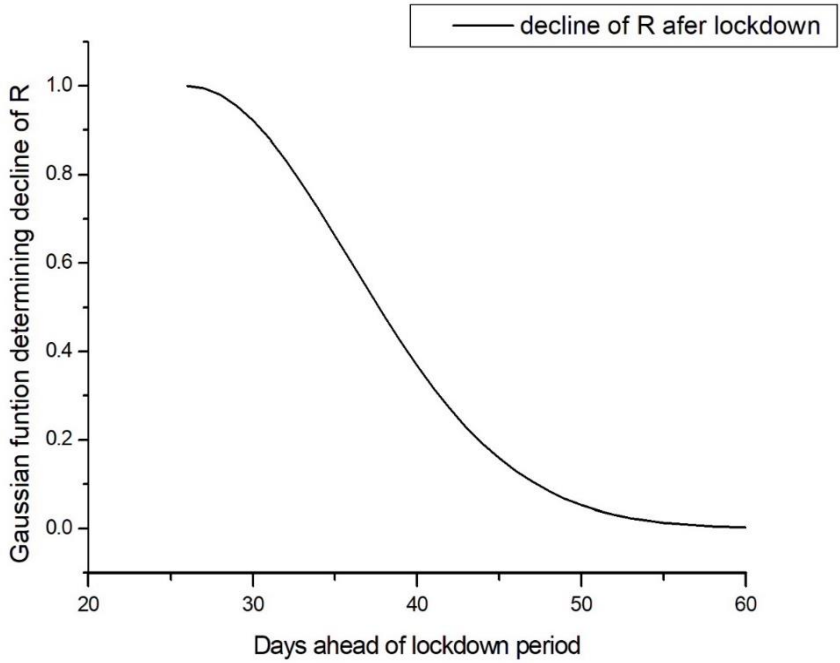

**Fig. 1. Assumed time dependence of R after lockdown plotted using τ₂=14 days as an example**

Finally, we arrive at the following eq for evolution of $N_e$

$$N_e(t) = N_e(0)e^{(R_0 \int_0^t e^{-\alpha(t-t_0)^2} dt - 1/\tau)} \qquad (7)$$

Further, splitting in time domain when $t \leq t_0$,

$$N_e(t) = N_e(0)e^{R_0 t} \qquad (8)$$

And for $t > t_0$,

$$N_e(t) = N_e(0)e^{R_0 t_0} e^{(R_0 \int_{t_0}^t e^{-\alpha(t-t_0)^2} dt - 1/\tau)} \qquad (9)$$

Which is solved as,

$$N_e(t) = N_e(0)e^{R_0 t_0} e^{\left(R_0 \sqrt{\frac{\pi}{\alpha}} \frac{1}{2} \text{erf}(\sqrt{\alpha}(t-t_0)) - (t-t_0)/\tau\right)} \qquad (10)$$

Therefore Eqs. 8 and 10 are the end results in terms of mean life time. The constants $N_e(0)$ and $R_0$ are obtainable from the observed data points of the infected daily data.

**RESULTS AND DISCUSSION**

The time τ₀ has been defined as announced by the Government of India on 24$^{th}$ March and taken to be 25 days. The whole calculation is based on times $\tau_0, \tau_1$ and $, \tau_2$. $N_e(0)$ and R₀ as fitted to $N_e e^{R_0 t}$ of all available from the data as available upto 16th April 2020, as shown in Fig.2, and as fitted ( 8.52 and 1/6, respectively). It is important to observe that active cases which differ from infected by the number recovered or died, is not significantly lowered. The fit of this curve is dictated by 1/6.36. The table I represents various input parameters used.

Table 1 (a) some decay times chosen

| Parameters chosen | Times in days |
|---|---|
| $\tau_0$ beginning of lock down (days) | 25+5* |
| $\tau_1$ recovery period of quarantine | 50 |
| $\tau_2$ relaxation time of $R_0$ under lockdown | 25 |

∗The 5 days period was added due to an event after lockdown referred to as Merkaz event

Table 1(b) The observed data exponential fit parameters

| Exponential fit parameters to infected cases $N_e(t) = N_e(0)e^{R_0 t}$ | Exponential fit parameters To only active cases $N_a(t) = N_a(0)e^{R_1 t}$ | Exponential fit parameters To dead cases $N_d(t) = N_d(0)e^{R_2 t}$ |
|---|---|---|
| $N_e(0) = 8.52$ | $N_0(0)=10.2$ | $N_d=0.116$ |
| $R_0= 1/6$ | $R_1=1/ 6.36$ | $R_3=1/5.28$ |

In Fig.2, the data as available[16] has been presented. This has been fitted to an exponential growth in Fig.3. Fig 4 presents a fit of the active data. The death count fit is given in Fig 5. Finally, the model based predicted curves are presented in Fig. 6.

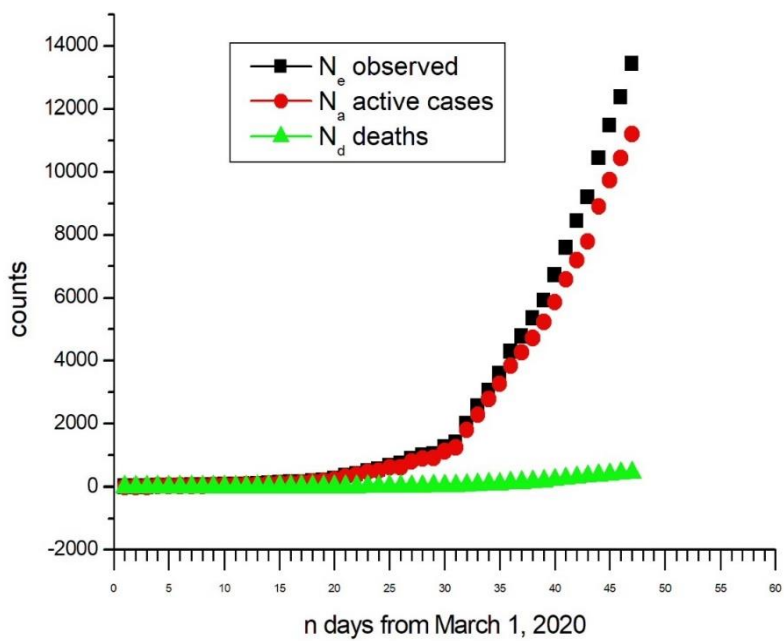

Fig. 2. The data of infected COVID-19 persons as shown in black squares. The active number of cases is differs from the total cases by recovered or dead cases.

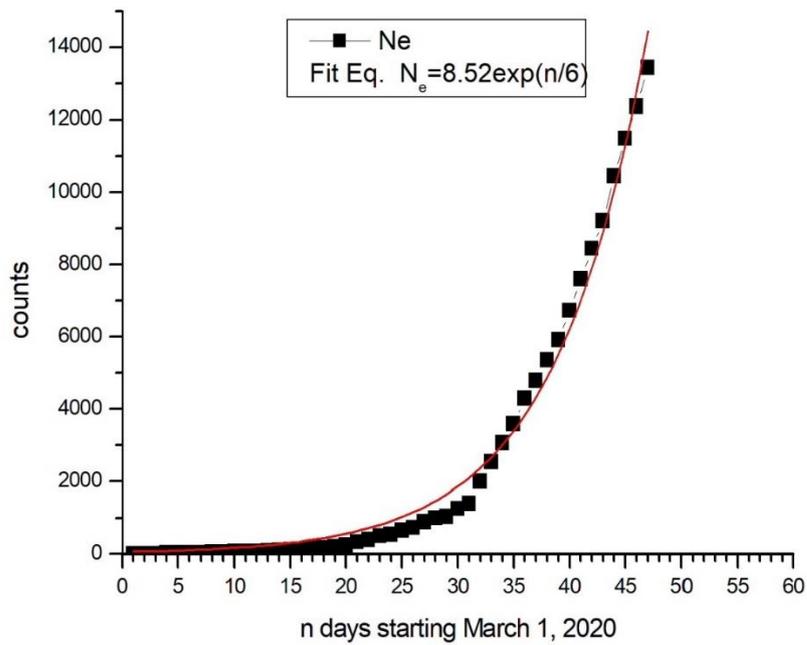

**Fig. 3. The fit to the data of infected COVID-19 patients in India.**

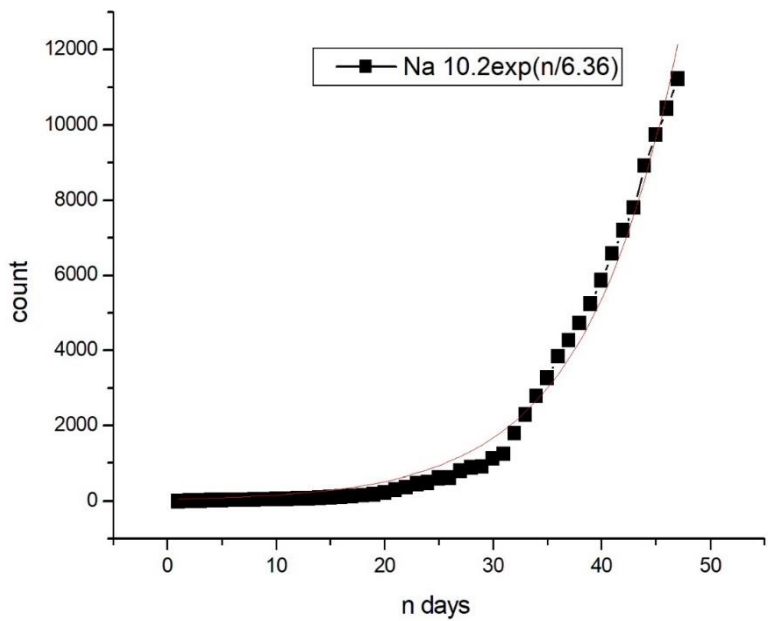

**Fig.4. The data and fit for the active number. The growth rate here $R_0^a$ =1/6.36 as compared to 1/6 for total number as shown in Fig. 2. This deviation helps us in fixing recovery mean life time τ$_1$.**

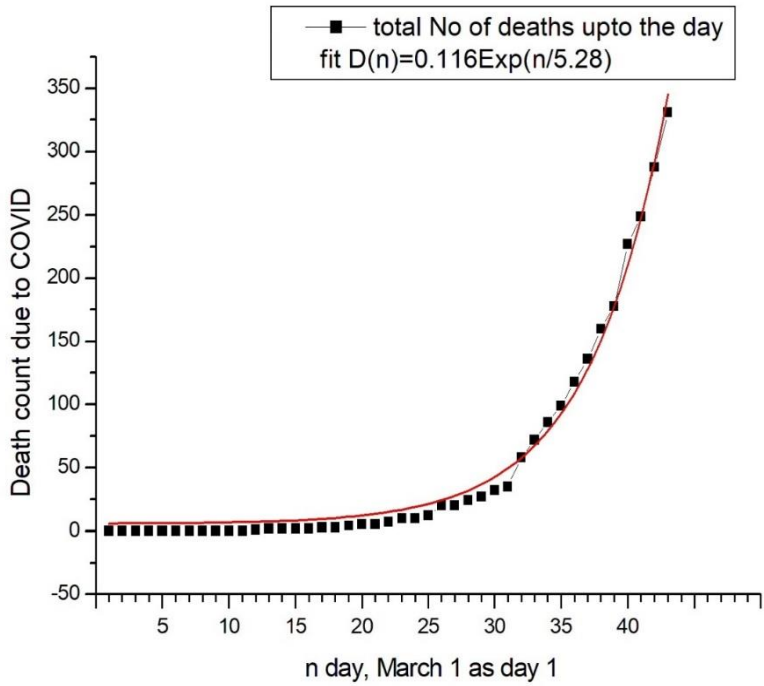

**Fig. 5. The Evolution of the count of deaths and its fit.**

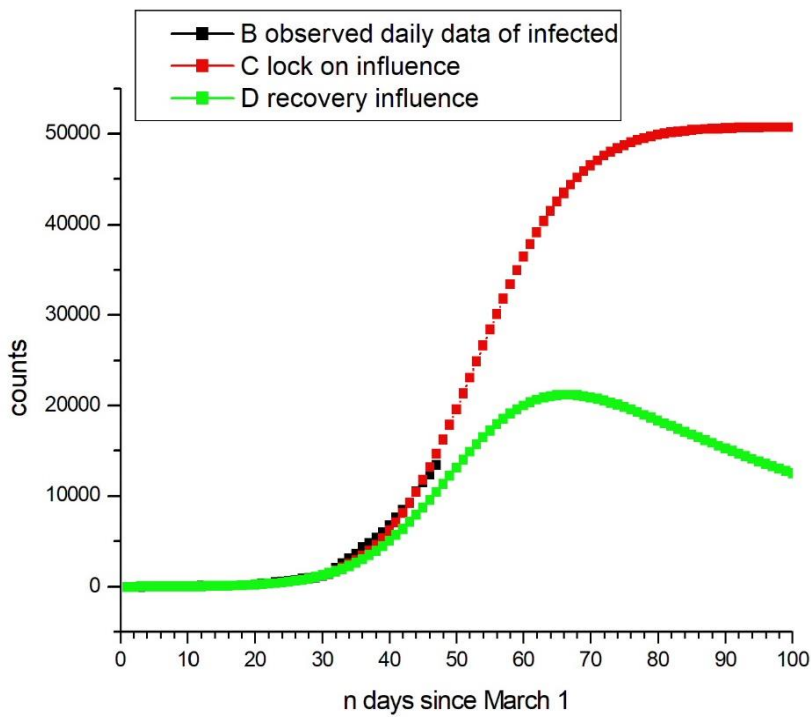

**Fig.6. The evolution of infected coronavirus persons as observed upto April 16, 2020 (black squares, and $N_e$ as modelled without incorporating decay (in red) and with survival recovery corrected through mean half life ( in green).**

As can be easily observed from Fig. 6, the observed data fits excellently to an exponential growth but with a small rate of increase. The fit has been incorporated into an expected evolution of infected number assuming a lockdown initiation. As R tends to zero, the number saturates. Again, when recovery is introduced in mean half life of 50 days, we get a peaked curve which may be a true picture, provided no new factors emerge. The initial recovery mean life considered was 14 days which turned out to be too quick and did not match with the recovery fit as one compares fit parameters of Ne and $N_a$.

It may be noted that if we assume that the number of infected $N_e$ is underquoted as the tests are not conducted thoroughly, we can update these by finding an updated $\widetilde{N_e}$ which has been grown from the death data, we get a significantly enhanced data, as plotted in Fig. 7.

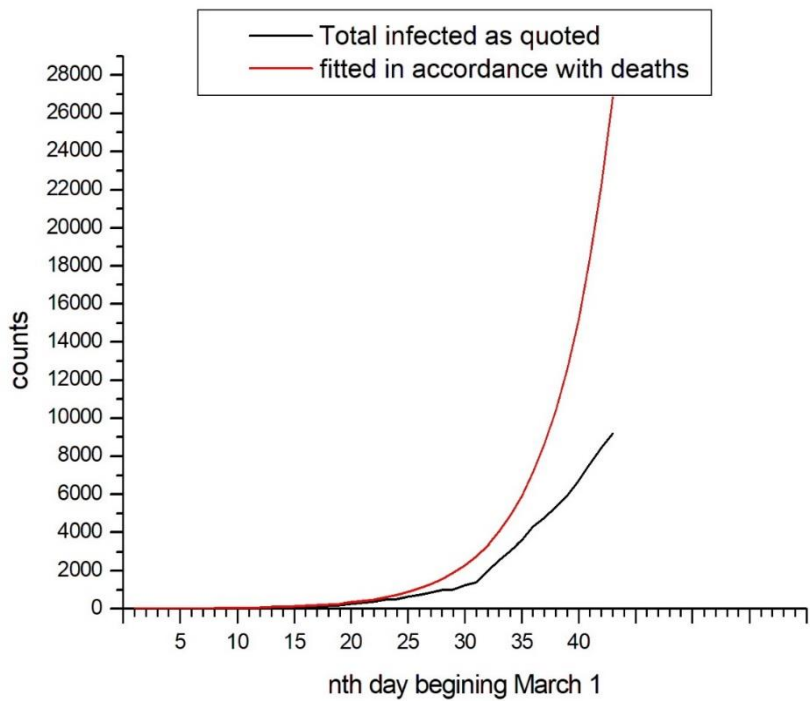

Fig. 7, $\widetilde{N}_e$ -a normalized $N_e$ as raised from the number of deaths due to COVID-19

Using this data will raise the data in Fig.7 by a factor of 3, approximately.

In conclusion, the aim of this paper is to give new ideas to model such cases of infected disease evolution. There may be possible objections to the choice of breadth of R after lockdown. There are also additional factors which develop as some hotspots emerging. A more realistic model needs to address these issues. However a new way to establish the data based theoretical model gives a new idea with great insight into how the factors are controlled and should be applied in most such studies instead of simulations where the contact with the processes is lost.

References:

1. Eili Klein et al, CDDEP report (March 24, 2020) https://cddep.org/wp-content/uploads/2020/03/covid19.indiasim.March23-2-eK.pdf
2. 2. Perc Matjaz, Miksić Nina Gorišek, Slavinec Mitja, and Stožer Andraž   Front. Phys., 08 April 2020 | https://doi.org/10.3389/fphy.2020.00127
3. Liu Y, Gayle AA, Wilder-Smith A, Rocklöv J. The reproductive number of COVID-19 is higher compared to SARS coronavirus. *J Travel Med.* (2020) **27**:taaa021. doi: 10.1093/jtm/taaa021
4. Remuzzi A, Remuzzi G. COVID-19 and Italy: what next? *Lancet*. (2020). doi: 10.1016/S0140-6736(20)30627-9.
5. Li Q, Guan X, Wu P, Wang X, Zhou L, Tong Y, et al. Early transmission dynamics in Wuhan, China, of novel coronavirus–infected pneumonia. *N Engl J Med.* (2020) **382**:1199–207. doi: 10.1056/NEJMoa2001316
6. Dong E, Du H, Gardner L. An interactive web-based dashboard to track COVID-19 in real time. *Lancet Infect Dis.* (2020). doi: 10.1016/S1473-3099(20)30120-1.
7. Zhao S, Lin Q, Ran J, Musa SS, Yang G, Wang W, et al. Preliminary estimation of the basic reproduction number of novel coronavirus (2019-nCoV) in China, from 2019 to 2020: a data-driven analysis in the early phase of the outbreak. *Int J Infect Dis.* (2020) **92**:214–7. doi: 10.1016/j.ijid.2020.01.050
8. Lai A, Bergna A, Acciarri C, Galli M, Zehender G. Early phylogenetic estimate of the effective reproduction number of SARS-CoV-2. *J Med Virol.* (2020). doi: 10.1002/jmv.25723.
9. Zhou T, Liu Q, Yang Z, Liao J, Yang K, Bai W, et al. Preliminary prediction of the basic reproduction number of the Wuhan novel coronavirus 2019-nCoV. *J Evid Based Med.* (2020) **13**:3–7. doi: 10.1111/jebm.12376
10. Ippolito G, Hui DS, Ntoumi F, Maeurer M, Zumla A. Toning down the 2019-nCoV media hype – and restoring hope. *Lancet Respir Med.* (2020) **8**:230–1. doi: 10.1016/S2213-2600(20)30070-9
11. Pastor-Satorras R, Castellano C, Van Mieghem P, Vespignani A. Epidemic processes in complex networks. *Rev Mod Phys.* (2015) **87**:925. doi: 10.1103/RevModPhys.87.925
12. Boccaletti S, Latora V, Moreno Y, Chavez M, Hwang D. Complex networks: structure and dynamics. *Phys Rep.* (2006) **424**:175–308. doi: 10.1016/j.physrep.2005.10.009
13. Holme P, Saramäki J. Temporal networks. *Phys Rep.* (2012) **519**:97–125. doi: 10.1016/j.physrep.2012.03.001

14. Boccaletti S, Bianconi G, Criado R, del Genio C, Gómez-Gardeñes J, Romance M, et al. The structure and dynamics of multilayer networks. *Phys Rep.* (2014) **544**:1–122. doi: 10.1016/j.physrep.2014.07.001
15. Wang Z, Bauch CT, Bhattacharyya S, d'Onofrio A, Manfredi P, Perc M, et al. Statistical physics of vaccination. *Phys Rep.* (2016) **664**:1–113. doi: 10.1016/j.physrep.2016.10.006
16. The data is publicly available
    https://www.worldometers.info/coronavirus/country/india/